\documentstyle[prd,aps,preprint,tighten,floats]{revtex}
\begin{document}
\draft
\preprint{UPR-658-T}
\date{May 1995}
\title{Four-dimensional supersymmetric dyonic
black holes in eleven-dimensional supergravity}
\author{Mirjam Cveti\v c
\thanks{E-mail address: cvetic@cvetic.hep.upenn.edu}
and Donam Youm
\thanks{E-mail address: youm@cvetic.hep.upenn.edu}}
\address{Physics Department \\
         University of Pennsylvania, Philadelphia PA 19104-6396}
\maketitle
\begin{abstract}
{A class of 4-dimensional supersymmetric dyonic black hole solutions
that arise in an effective 11-dimensional supergravity compactified on
a 7-torus is presented.  We give the explicit form of dyonic solutions
with diagonal internal metric, associated with the Kaluza-Klein
sector as well as the three-form field, and relate them to a class
of solutions with a general internal metric by imposing a subset of
$SO(7)\subset E_7$ transformations.  We also give the field transformations
which relate the above configurations to 4-dimensional ground state
configurations of Ramond-Ramond and Neveu-Schwarz-Neveu-Schwarz sector
of type-IIA strings on a 6-torus.}
\end{abstract}

\section{Introduction}

In theories that attempt to unify gravity with other forces of nature,
in particular, effective theories from superstrings, the non-trivial
configurations, {\it i.e.}, topological defects as well as black holes
(BH's), provide an important testing ground to address the role of gravity
in such theories.  Configurations which saturate the Bogomol'nyi bound on
their energy (ADM mass) correspond to the ground state configurations within
its class.  Supersymmetric embedding of such configurations ensures that
they are invariant under (constrained) supersymmetry transformations,
{\it i.e.}, they satisfy the corresponding Killing spinor equations.  For
the above reasons, one refers to such configurations as supersymmetric, and
in the case of spherically symmetric configurations, as Bogomol'nyi-Prasad-
Sommerfeld (BPS) saturated states.

BPS saturated states within effective (super)gravity theories are of
special interest, since they shed light on the nature of non-trivial
ground states in such theories.  In view of recently conjectured duality
\cite{WITTEN,UDUAL,TOWN2} between strongly coupled 10-d type-IIA superstring
theory and 11-d supergravity (SG) theory compactified on a circle as well as
related dualities in other dimensions \cite{SEN2,STRO}
\footnote{For a related recent works see Refs.
\cite{BARS,GATES,BERG,TOWN,STRO2,STRO3,VAFA} and references therein.},
{\it e.g.}, the strong-weak coupling limits of type-IIA string theory
compactified on $K_3$ surface, and the 11-d SG and heterotic string theory
compactified on a torus, it is important to have a systematic and explicit
construction of {\it all} the supersymmetric configurations of the
corresponding effective SG theories in different dimensions.
Such a systematic study would, in turn, provide a ground work for testing
explicitly the conjectured duality symmetries between the corresponding
effective (weakly coupled) SG theories and the corresponding strongly
coupled string theories in different dimensions, at least at the level of
the light (ground state) spectrum of states.
The proposed  construction of all the BPS saturated states is a monumental
task, however, some of its aspects have been addressed for certain types of
supersymmetric configurations in certain dimensions\cite{SEN2,STRO,BERG,TOWN}.

In this paper, we would like to single out a specific aspect of the proposed
study.  Namely, we would like to set up an explicit construction of 4-d BPS
saturated BH solutions of 11-d SG compactified on a 7-torus
\footnote{Compactification on a 7-torus, {\it i.e.}, the internal isometry
group is Abelian, provides the simplest possible compactification,
and thus the first one to be addressed.}.
These configurations would, in turn, allow for the testing of the
conjectured duality between these states and the light spectrum of the
corresponding strongly coupled string theory(ies) in 4-d.

BPS saturated states, corresponding to four-dimensional (4-d), spherically
symmetric, static BH solutions, have masses which are related to the electric
and magnetic charges of the BH.  With the electric and magnetic charges
quantized, masses of such BH's are related to the multiples of the elementary
electric and magnetic charges.  A subset of 4-d supersymmetric BH's within
different sectors of effective 4-d SG theories has been already addressed
\footnote{See Refs.\cite{MAX,GIBB,KHURI,HOST,HARVEY,COS,GIVE,SEN1,BANKS,GAUNT}
and references therein.}.
However, results were primarily obtained in special cases with either
non-zero electric or non-zero magnetic charges. In addition, only a subset
of scalar fields was turned on
\footnote{Recently, a general class of electrically charged, rotating
BH solutions in the heterotic string theory compactified on a 6-torus
has been constructed \cite{SEN1}, and a procedure to construct the
corresponding solutions with a general electric and magnetic charge
configurations has been spelled out.  Supersymmetric limit of the latter
configurations, however, has not been addressed, yet.}.

Recently, progress has been made in finding the explicit form of all the
4-d supersymmetric as well as all the non-extreme BH's in Abelian $(4+n)$-d
Kaluza-Klein (KK) theory.  Those are BH's with the $U(1)$ gauge fields and
scalar fields which originate from a $(4+n)$-d metric turned on.
Supersymmetric BH's consist of $n$ electric charges $\vec{\cal Q}
\equiv (Q_1,...,Q_n)$ and $n$ magnetic charges $\vec{\cal P} \equiv
(P_1,...,P_n)$ which are subject to the constraint $\vec{\cal P}\cdot
\vec{\cal Q}=0$ \cite{ADD}. The generating solutions are supersymmetric
$U(1)_M\times U(1)_E$ BH's \cite{SUPER}, {\it i.e.}, dyonic BH's with one
electric and one magnetic charges arising from different $U(1)$ factors,
which correspond to supersymmetric BH's with a diagonal internal metric
Ansatz.  All the supersymmetric BH's are obtained by performing the
$SO(n)/SO(n-2)$ rotations, which do not affect the 4-d space-time metric
and the volume of the internal space, on the generating solution, {\it i.e.},
the supersymmetric $U(1)_M \times U(1)_E$ solution.  The explicit form for
all the 4-d, Abelian, static, spherically KK BH's is obtained \cite{ALLKKBH}
by performing two $SO(1,1)$ boosts on the non-extreme $U(1)_M \times U(1)_E$
KK BH and supplementing it by $SO(n)/SO(n-2)$ transformations.

The aim of this paper is a few fold.  First, we provide an explicit
construction of 4-d supersymmetric BH configurations which arise from
two different sectors of 11-d SG theory compactified on a 7-torus.
The first one is the configurations whose charges arise from $U(1)$
gauge fields associated with the 11-d metric (KK BH's).  Abelian BH's
in the $(4+n)$-d KK theory with $n=7$ constitute this subset of BH
solutions.  The second class of configurations are those whose charges
arise from the compactification of the three-form field $A^{(11)}_{MNP}$
of 11-d SG.  The explicit general construction of the latter
configurations, with the scalar fields of the internal metric and the
volume of the internal space turned on, constitutes a major part of
this paper.  In addition, we address the symmetry structure of such
configurations and the procedure to obtain an explicit form of all such
ground state configurations.  The two classes of such solutions provide
the initial building blocks to construct all the supersymmetric BH's of
11-d SG on a 7-torus.

Another aim is to address the connection between the strongly coupled
states of type-IIA string theory and those of weakly coupled 11-d SG.
In particular, we find the explicit field transformations between
4-d solutions of 11-d SG and Ramond-Ramond (R-R) as well as Neveu-Schwarz-
Neveu-Schwarz (NS-NS) sector of type-IIA superstring theory compactified
on a 6-torus.

The paper is organized as follows.  In chapter II, we summarize the
properties of 11-d SG theory and its compactification down to 4-d.  In
chapter III, we study two classes of Abelian charged BH solutions, each
of which are associated with KK gauge fields and 3-form $U(1)$ gauge fields,
respectively.  In chapter III, we discuss the connection between the 11-d
SG and type-IIA, as well as heterotic superstring theory, and comment on
the strong-weak coupling behavior among such theories. In chapter IV,
conclusions are given.

\section{Eleven-dimensional supergravity and its compactification down to
four-dimensions}

In this section, we summarize the particle content and the effective
Lagrangian density of 11-d supergravity (SG) compactified down to 4-d
on a 7-torus.  The field content of the $N$=1, d=11 SG is the
Elfbein $E^{(11)\, A}_M$, gravitino $\psi^{(11)}_M$, and the 3-form
field $A^{(11)}_{MNP}$.  The bosonic Lagrangian density is given by
\cite{ELE}
\begin{equation}
{\cal L} = -{1\over 4}E^{(11)}[{\cal R}^{(11)} + {1\over {12}}
F^{(11)}_{MNPQ}F^{(11)\ MNPQ} - {8\over {12^4}}
\varepsilon^{M_1 \cdots M_{11}}F_{M_1 \cdots M_4}F_{M_5 \cdots M_8}
A_{M_9 M_{10} M_{11}}],
\label{action11d}
\end{equation}
where $E^{(11)} \equiv {\rm det}\, E^{(11)\ A}_M$, ${\cal R}^{(11)}$
is the Ricci scalar defined in terms of the Elfbein, and
$F^{(11)}_{MNPQ} \equiv 4\partial_{[M}A^{(11)}_{NPQ]}$ is the field
strength associated with the 3-form field $A^{(11)}_{MNP}$.
The supersymmetry transformation of the gravitino field $\psi^{(11)}_M$
in the bosonic background is given by
\begin{equation}
\delta \psi^{(11)}_M = D_M\, \varepsilon +{i\over 144} (\Gamma^{NPQR}_
{\ \ \ \ \ M} - 8\Gamma^{PQR}\delta^N_M)F_{NPQR}\,\varepsilon ,
\label{sstran11d}
\end{equation}
where $D_M\,\varepsilon = (\partial_M + {1\over 4}\Omega_{MAB}
\Gamma^{AB})\,\varepsilon$ is the gravitational covariant derivative
on the spinor $\varepsilon$, and $\Omega_{ABC} \equiv -\tilde{\Omega}_
{AB,C} + \tilde{\Omega}_{BC,A} - \tilde{\Omega}_{CA,B}$
($\tilde{\Omega}_{AB,C} \equiv E^{(11)\,M}_{[A}E^{(11)\,N}_{B]}
\partial_N E^{(11)}_{MC}$) is the spin connection defined in terms of
the Elfbein.  Our convention for the metric signature is
($+--\cdots -$).  For the space-time vector index convention,
the characters ($A,B,...$) and  ($M,N,...$) denote flat and curved
indices, respectively.

The dimensional reduction of 11-d theory down to 4-d is achieved by
taking the KK Ansatz for the Elfbein and a consistent truncation of the
other 11-d fields \cite{CRE}.  With the internal space being a 7-torus
$T^7$ ,{\it i.e.}, the internal isometry group is Abelian, all the
fields are independent of the internal coordinates in the zero mode
approximation.  One can use the off-diagonal part of local Lorentz
$SO(1,10)$ invariance to put the Elfbein into the following triangular
form:
\begin{equation}
E^{(11)\,A}_M = \left ( \matrix{e^{-{\varphi \over 2}} e^{\alpha}_{\mu} &
B^i_{\mu} e^a_i \cr 0 & e^a_i} \right ),
\label{elfbein}
\end{equation}
where $\varphi \equiv {\rm ln}\,{\rm det}\,e^a_i$, and $B^i_{\mu}$
($i=1,...,7$) are KK Abelian gauge fields.  Here, we use
the greek letters ($\alpha, \beta ,\cdots$) [($\mu ,\nu ,\cdots$)] for the
4-d space-time flat [curved] indices and the latin letters ($a,b,\cdots$)
[($i,j,\cdots$)] for the internal flat [curved] space indices.
The 3-form field $A^{(11)}_{MNP}$ is truncated into the following three
types of 4-d fields: 35 pseudo-scalar fields $A_{ijk}$, 21 pseudo-vector
fields $A_{\mu\,ij}$ and 7 two-form fields $A_{\mu\nu\,i}$. The two-form
fields $A_{\mu\nu\,i}$ are equivalent to (axionic) scalar fields
$\varphi^i$ after making duality transformation.  In order to ensure that
the fields $A_{\mu\,ij}$ and $A_{\mu\nu\,i}$ are scalars under the internal
coordinate transformation $x^i \to x^{\prime\ i} = x^i + \xi^i$, and
transform as $U(1)$ gauge fields under the gauge transformation
$\delta A^{(11)}_{MNP} = \partial_M \zeta_{NP} + \partial_N \zeta_{PM} +
\partial_P \zeta_{MN}$, one has to define new canonical 4-d fields in the
following way:
\begin{equation}
A^{\prime}_{\mu\,ij} \equiv A_{\mu\,ij} - B^k_{\mu} A_{kij},\ \ \ \
A^{\prime}_{\mu\nu\,i} \equiv A_{\mu\nu\,i} - B^j_{\mu}A_{j\nu\,i}
-B^j_{\nu}A_{\mu\,ji} + B^j_{\mu} B^k_{\nu} A_{jki}.
\label{cantensor}
\end{equation}
Then, the bosonic action (\ref{action11d}) reduces to the following
effective 4-d action:
\begin{equation}
{\cal L} = -{1\over 4}e[{\cal R} - {1\over 2}\partial_{\mu} \varphi
\partial^{\mu} \varphi +{1\over 4}\partial_{\mu} g_{ij} \partial^{\mu}
g^{ij} - {1\over 4} e^{\varphi}g_{ij}G^i_{\mu\nu}G^{j\,\mu\nu}
+{1\over 2}e^{\varphi}g^{ik}g^{jl}F^4_{\mu\nu\,ij}
F^{4\,\mu\nu}_{\ \ \ \ kl} + \cdots ],
\label{action4d}
\end{equation}
where $e \equiv {\rm det}\,e^\alpha_\mu$, the Ricci scalar ${\cal R}$ is
defined in terms of the Einstein-frame 4-d metric $g_{\mu\nu} = \eta_
{\alpha\beta}e^{\alpha}_{\mu}e^{\beta}_{\nu}$, $G^i_{\mu\nu} \equiv
\partial_{\mu} B^i_{\nu} - \partial_{\nu}B^i_{\mu}$, $F^4_{\mu\nu\,ij}
\equiv F^{\prime}_{\mu\nu\,ij} + G^k_{\mu\nu}A_{ijk}$, and the dots
($\cdots$) denotes the terms involving the pseudo-scalars $A_{ijk}$ and
the two-form fields $A_{\mu\nu\,i}$.  Here, $g_{ij} \equiv \eta_{ab}e^a_i
e^b_j = -e^a_i e^a_j$ is the internal metric and the curved space indices
($i,j,...$) are raised by $g^{ij}$
\footnote{The dilaton field $\varphi$ and the internal metric $g_{ij}$
in Eq. (\ref{action4d}) are related to the dilaton field $\varphi$ and
the unimodular part of the internal metric $\rho_{ij}$, used in
Ref. \cite{SUPER} as $\varphi \rightarrow {\sqrt{7} \over 3}
{\varphi}\ \ , \ \ g_{ij} \rightarrow - e^{{2 \over {3\sqrt{7}}}
{\varphi}}\rho_{ij}$.}.
The 4-d effective action (\ref{action4d}) is manifestly invariant
under the $SL(7,\Re)$ target space transformation:
\begin{equation}
g_{ij} \to U_{ik} g_{kl} U_{jl},\ \ \ \
G^i_{\ \mu\nu} \to (U^{-1})_{ik} G^k_{\ \mu\nu} , \ \ \ \
F^4_{\mu\nu\,ij} \to (U^{-1})_{ki} (U^{-1})_{lj} F^4_{\mu\nu\,kl},
\label{GL7}
\end{equation}
and the dilaton $\varphi$ and the 4-d metric $g_{\mu\nu}$ remain intact,
where $U \in SL(7,\Re)$.

The $SL(7,\Re)$ target space symmetry can be enlarged to the global
$SL(8,\Re)$ symmetry by realizing that 7 scalars $\varphi^i$, which are
equivalent to $F_{\mu\nu\rho\,i}$ through the duality transformation,
$F^{\mu\nu\rho}_{\ \ \ i} \sim (\sqrt{-g})^{-1}e^{-2\varphi}g_{ij}
\varepsilon^{\mu\nu\rho\sigma}\partial_{\sigma}\varphi^j$, are unified
with the Siebenbein $e^i_a$ to form a matrix  parameterizing  $SL(8,\Re)$
\cite{CRE}.  In this case, 7 KK gauge fields $B^k_{\mu}$ and 21
``magnetic'' duals $B^{ij}_{\mu}$ of $A_{\mu\,ij}$ form canonical gauge
fields of 28 $U(1)$ groups.  With a further inclusion of 35 pseudo-scalar
fields $A_{ijk}$, the $SL(8,\Re)$ group is enlarged to the exceptional
group $E_7$, in which case 28 $U(1)$ gauge fields and their 28 dual
fields form the $\bf 56$ fundamental representation of $E_7$.

The elements of the above enlarged symmetry groups can be used to provide
transformations on existing 4-d solutions, thus generating a family of
solutions with the {\it same} 4-d space-time $g_{\mu\nu}$ and dilaton field
$\varphi$, however, with transformed dyonic charges and other scalar fields.

In this paper, we find a class of 4-d, supersymmetric, Abelian solutions
where all the scalar fields, except the dilaton and the diagonal components
of the internal metric, are set to zero.  We primarily concentrate on a
subset of $SO(7)\subset SL(7,\Re)$ transformation on these basis solutions
in order to construct the most general family of solutions with scalar fields
other than the internal metric $g_{ij}$ and the dilaton $\varphi$ turned off.
The ultimate goal, however, is to find the generating solutions which,
supplemented by a subset of $E_7$ transformations, would generate {\it all}
the supersymmetric BH solutions with all the scalar fields turned on
\footnote{Within the context of Abelian KK BH's, such a program has been
completed.  $SO(n)/SO(n-2)$ transformations on the $U(1)_M \times U(1)_E$
supersymmetric solutions \cite{SUPER} generate the most general
supersymmetric static BH's \cite{ADD} in KK theory.  On the other hand,
in order to generate the most general non-extreme solutions \cite{ALLKKBH},
one has to employ a subset of a larger symmetry, {\it i.e.}, a subset
of $SO(2,n)\subset SL(n+2,\Re)$ transformations, which correspond to the
symmetry transformations of effective 3-d action of stationary solutions.}.
We conjecture that {\it all} the supersymmetric BH's of 11-d SG are
generated by imposing a subset of $E_7$  transformations
\footnote{In the quantum version with charges quantized in multiples of
the elementary electric and magnetic charges, $E_7$ would be broken
down to the corresponding discrete subgroup.}
(of the effective 4-d action) on only {\it one} type of the generating
solution. Such a generating solution would reduce to two separate classes,
which are discussed in the next chapter, by taking special limits of charge
configurations.  On the other hand, in order to generate all the
non-extreme BH's, one would have to employ a larger symmetry of the
corresponding 3-d effective action for stationary solutions.  We speculate
that such an enlarged symmetry might be $E_8$.

\section{Four-dimensional, supersymmetric, static, spherically symmetric
solutions of eleven-dimensional supergravity}

In this section, we study supersymmetric, static, spherically symmetric
configurations arising from two sectors of effective 4-d theory
(\ref{action4d}):
\begin{itemize}
\item
The first sector corresponds to configurations with non-zero electric
and magnetic charges arising only from gauge fields of Abelian internal
isometry group, {\it i.e.}, KK gauge fields $B_{\mu}^i\neq 0$.
We will refer to this class of solutions as supersymmetric ``KK BH's''.
\item
The second sector corresponds to configurations with non-zero electric
and magnetic charges only from the Abelian gauge fields $A_{\mu\,ij}$
associated with the 3-form field $A^{(11)}_{MNP}$. We will refer to this
class of solutions as supersymmetric ``3-form BH's''.
\end{itemize}
Within each  class of solutions, we shall obtain the most general solution,
with non-zero internal metric $g_{ij}$ and dilaton field $\varphi$, while
the other scalar fields are turned off.

The spherically symmetric Ansatz for the 4-d space-time metric
\footnote{The moving frame is then defined in terms of the Vierbein
of the following form:
\begin{eqnarray}
e^{\hat{t}}_t = \lambda^{1\over 2},\ \ \
e^{\hat{\theta}}_{\theta} = R^{1\over 2}, \ \ \
e^{\hat{\phi}}_{\phi} = R^{1\over 2}{\rm sin}\theta ,\ \ \
e^{\hat{r}}_r = \lambda^{-{1\over 2}}, \nonumber
\end{eqnarray}
which yields the metric $g_{\mu\nu} = \eta_{\alpha\beta} e^{\alpha}_{\mu}
e^{\beta}_{\nu}$ defined above.  Here, $\alpha , \beta  = \hat{t},
\hat{\theta}, \hat{\phi}, \hat{r}$ are flat indices and the flat
space-time gamma matrices are ordered in the same manner, {\it i.e.},
$\gamma^{\hat{t}} = \gamma^0,...,\gamma^{\hat{r}} = \gamma^3$.}
is chosen to be
\begin{equation}
g_{\mu\nu}{\rm d}x^{\mu}{\rm d}x^{\nu} = \lambda (r){\rm d}t^2
- \lambda^{-1} (r) {\rm d}r^2 - R(r)({\rm d}\theta^2 +{\rm sin}^2 \theta
{\rm d}\phi^2),
\label{met4d}
\end{equation}
and the scalar fields depend on the radial coordinate $r$ only.
For a particular $U(1)$ gauge field $A_\mu$, the non-zero components,
compatible with the spherical symmetry, are given in the polar coordinate by
\begin{equation}
A_{\phi} = P(1-{\rm cos}\theta), \ \ \ \ \ A_t = \psi(r),
\label{vecpot}
\end{equation}
where $E(r) = -\partial_r \psi(r) \sim {Q \over {r^2}}$ ($r \to \infty$),
and $P$ and $Q$ are the physical magnetic and electric charges.  Here,
the expression for an electric field, {\it i.e.}, the $(t,r)$ component
of the $U(1)$ field strength, in terms of the scalar fields and the 4-d
metric components are obtained from the Gauss's law derived from the
Lagrangian density (\ref{action4d}).  The expressions for two types of
electric field strengths are given in the following form:
\begin{eqnarray}
\nabla_r(e^{\varphi}g_{ij}G^{j\,rt})=0 \ \ \ &\Longrightarrow& \ \ \
G^i_{tr} = {{g^{ij}\tilde{Q}_j} \over {Re^{\varphi}}},
\nonumber \\
\nabla_r(e^{\varphi}g^{ik}g^{jl}F^{rt}_{kl})=0 \ \ \
&\Longrightarrow& \ \ \  F_{tr\,ij} = {{g_{ik}g_{jl}\tilde
{Q}^{kl}} \over {Re^{\varphi}}} \ \ ({\rm with}\ B^i_{\mu} = 0),
\label{elec}
\end{eqnarray}
where the physical electric charges are given by $Q^i = e^{-\varphi_{\infty}}
g^{ij}_{\infty}\tilde{Q}_j$ and $Q_{ij} = e^{-\varphi_{\infty}}
g_{ik\,\infty}g_{jl\,\infty}\tilde{Q}^{kl}$.

\subsection{Kaluza-Klein Black Hole Solutions}

The first class of solutions corresponds to supersymmetric KK BH's,
{\it i.e.}, $U(1)$ gauge fields are associated with the isometry group
of the internal space.  In this case, the gauge fields arising from the
3-form field are turned off, along with all the scalar fields,
except the internal metric fields $g_{ij}$ and the dilaton $\varphi$.
Then, the Lagrangian density (\ref{action4d}) reduces to the 4-d
effective Lagrangian density of 11-d KK theory compactified on a 7-torus.
The corresponding action has a manifest invariance under $SO(7)\subset
SL(7,\Re)$ rotations (see Eq.(\ref{GL7})) with the internal metric
coefficients $g_{ij}$ transforming as a symmetric $\bf 27$ representation
of $SO(7)$, and the $U(1)$ gauge fields $B_\mu^i$ as $\bf 7$ of $SO(7)$.

Since 4-d, Abelian, supersymmetric BH's of $(4+n)$-d KK theory have
been explicitly constructed in Refs. \cite{SUPER,ADD}, here, we summarize
the results for the special case of $n=7$.

With a diagonal internal metric Ansatz, supersymmetric spherically
symmetric configurations choose the vacuum where the isometry group of
the internal space is broken down to $U(1)_M \times U(1)_E$, {\it i.e.},
they correspond to a dyonic configuration with magnetic and electric
charges coming from different $U(1)$ gauge groups.  The most general
supersymmetric spherically symmetric configurations with a non-diagonal
internal metric $g_{ij}$ are obtained by imposing the $SO(7)/SO(5)$
rotations on the $U(1)_M \times U(1)_E$ solutions.  The charge vectors
$\vec{\cal P} \equiv (P_1,...,P_7)$ and $\vec{\cal Q} \equiv
(Q_1,...,Q_7)$ of this general class of supersymmetric solutions
are constrained by $\vec{\cal P} \cdot \vec{\cal Q} = 0$, thereby having
$2n - 1 = 13$ degrees of freedom.

Explicit supersymmetric $U(1)_M \times U(1)_E$ solutions of 11-d
SG with the $j$-$th$ gauge field magnetic and the $k$-$th$ gauge field
electric, and with a diagonal internal metric Ansatz $g_{ij}={\rm diag}
(g_{11},\cdots ,g_{77})$, are given by
\begin{eqnarray}
\lambda &=& {{r - |{\bf P}_{j\,\infty}| - |{\bf Q}_{ k\,\infty}|} \over
{(r - |{\bf P}_{ j\,\infty}|)^{1\over 2}(r - |{\bf Q}_{ k\,\infty}|)^
{1\over 2}}},
\nonumber \\
R &=& r^2 (1 - {{|{\bf P}_{j\,\infty}| +|{\bf Q}_{k\,\infty}|}\over r})
(1 - {{|{\bf P}_{j\,\infty}|} \over r})^{1 \over 2}
(1 - {{|{\bf Q}_{k\,\infty}|} \over r})^{1 \over 2},
\nonumber \\
e^{{2 }({\varphi} - {\varphi}_{\infty})} &=&
{{r - |{\bf P}_{j\,\infty}|} \over {r - |{\bf Q}_{k\,\infty}|}},
\nonumber \\
{{g_{ii}}\over {g_{ii\,\infty}}} &=&
1\ \ \ \ \ (i \neq j,k)\ \ ,
\nonumber \\
{{g_{jj}}\over{g_{jj\,\infty}}}&=& {{r - |{\bf P}_{j\,\infty}| -
|{\bf Q}_{k\,\infty}|} \over {(r - |{\bf Q}_{k\,\infty}|)}},
\nonumber \\
{g_{kk}\over g_{kk\,\infty}} &=& {{(r - |{\bf P}_{j\,\infty}|)}
\over {r - |{\bf P}_{j\,\infty}| - |{\bf Q}_{k\,\infty}|}},
\nonumber \\
a^{\bf m}_u (r) &=& a^{\bf m}_{u\,\infty}\left ({{r-|{\bf P}_{j\,\infty}|
-|{\bf Q}_{k\,\infty}|} \over {r-|{\bf P}_{j\,\infty}|}}\right )^{1\over 4},
\label{kksol}
\end{eqnarray}
where ${\bf P}_{j\infty} \equiv e^{{1\over 2}{\varphi}_{\infty}}
g^{1\over 2}_{jj\,\infty}P_j$ and ${\bf Q}_{k\infty} \equiv e^{{1\over 2}
{\varphi}_{\infty}}g^{1\over 2}_{kk\,\infty}Q_k$ are the ``screened''
magnetic and electric charges (here, $P_j$ and $Q_k$ are the physical
magnetic and electric charges of the $j$-$th$ and $k$-$th$ gauge fields).
The subscript $\infty$ refers to the asymptotic ($r\rightarrow \infty)$
value of the corresponding scalar field.  The ADM mass of the configuration
is given by $M = |{\bf P}_{j\,\infty}| + |{\bf Q}_{k\,\infty}|$.

For $P_j \neq 0$ and $Q_k \neq 0$, 4-d space-time has a null singularity,
finite temperature ($T_H = (4\pi)^{-1}|{\bf P}_{j\infty}{\bf Q}_{k\infty}|^
{-1/2}$) and zero entropy. If either of $P_j$ or $Q_k$ is set equal to zero,
the singularity becomes naked and the temperature diverges.

\subsection{Three-Form Black Hole Solutions}

The second class of solutions is static, spherically symmetric
configurations associated with the 21 Abelian pseudo-vector fields
$A_{\mu\,ij}$, arising from the 3-form fields $A^{(11)}_{MNP}$.
In this case, we set the KK gauge fields $B^i_{\mu}$ equal to zero,
as well as all the other scalar fields except those associated with
the internal metric $g_{ij}$ and the dilaton $\varphi$.  The
corresponding 4-d bosonic effective Lagrangian density is then of the
following from:
\begin{equation}
{\cal L} = -{1\over 4}e[{\cal R} - {1\over 2}\partial_{\mu} \varphi
\partial^{\mu} \varphi + {1\over 4} \partial_{\mu} g_{ij} \partial^{\mu}
g^{ij} + {1\over 2}e^{\varphi} g^{ik}g^{jl}F_{\mu\nu\,ij}
F^{\mu\nu}_{\ \ \,kl}],
\label{action3}
\end{equation}
where $F_{\mu\nu\,ij} \equiv \partial_{\mu}A_{\nu\,ij} - \partial_{\nu}
A_{\mu\,ij}$.  The bosonic action (\ref{action3}) has again a manifest
invariance under the $SO(7)\subset SL(7,\Re )$ rotations (see Eq.
(\ref{GL7})) with the internal metric coefficients $g_{ij}$ transforming
as a symmetric representation $\bf 27$ of $SO(7)$, and the $U(1)$ gauge
fields $A_{\mu\, ij}$ as an antisymmetric representation $\bf 21 $
of $SO(7)$.

\subsubsection{Killing Spinor Equations}

The supersymmetric solutions of the above action are invariant under the
gravitino transformations (\ref{sstran11d}), which for the above bosonic
field content reduce to the following form:
\begin{eqnarray}
\delta \psi_{\mu} &=& \partial_{\mu} \varepsilon + {1\over 4}
\omega_{\mu\beta\gamma} \gamma^{\beta\gamma} \varepsilon -
{1 \over 4}e^{\alpha}_{\mu}\eta_{\alpha[\beta}e^{\nu}_{\gamma ]}
\partial_{\nu} \varphi \gamma^{\beta\gamma} \varepsilon +
{1 \over 8}(e^l_b \partial_{\mu}e_{lc} - e^l_c
\partial_\mu e_{lb})\gamma^{bc} \varepsilon
\nonumber \\
& &+ {i \over {24}}e^{\varphi \over 2}
F_{\nu\rho\,ij}\gamma^{\nu\rho}_{\ \ \mu}\gamma^{ij} \varepsilon -
{i \over 6}e^{\varphi \over 2}F_{\mu\nu\,ij}
\gamma^{\nu}\gamma^{ij}\varepsilon , \nonumber \\
\delta \psi_k &=& -{1\over 4}e^{\varphi \over 2}(\partial_{\rho} e_{kb} +
e^c_k e^l_b \partial_{\rho} e_{lc})\gamma^{\rho 5}
\gamma^b \varepsilon +{i \over {24}}e^{\varphi}F_{\mu\nu\,ij}
\gamma^{\mu\nu 5}\gamma^{ij}_{\ \  k}\varepsilon - {i\over 6}
e^{\varphi}F_{\mu\nu\,kl}\gamma^{\mu\nu 5}\gamma^l\varepsilon ,
\label{sstran4d}
\end{eqnarray}
where $\omega_{\mu\beta\gamma}$ is the spin-connection defined in terms
of the Vierbein $e^{\alpha}_{\mu}$ and $[a\,\cdots\,b]$ denotes the
antisymmetrization of the corresponding indices.  For the 11-d gamma
matrices, which satisfy the $SO(10,1)$ Clifford algebra $\{ \Gamma^A,
\Gamma^B \} = 2\eta^{AB}$, we have chosen the following representation:
\begin{equation}
\Gamma^{\alpha} = \gamma^{\alpha} \otimes I , \ \ \ \ \ \
\Gamma^a = \gamma^5 \otimes \gamma^a ,
\label{gamma}
\end{equation}
where $\{ \gamma^{\alpha}, \gamma^{\beta} \} = 2\eta^{\alpha\beta}$,
$\{ \gamma^a , \gamma^b \} = -2\delta^{ab}$, $I$ is the $8 \times 8$
identity matrix and $\gamma^5 \equiv i\gamma^0 \gamma^1 \gamma^2
\gamma^3$.  The above representation (\ref{gamma}) is compatible
with the triangular gauge form (\ref{elfbein}) ($SO(10,1) \to SO(3,1)
\times SO(7)$) of the Elfbein.  The gamma matrices with more than one
index denote antisymmetric products of the corresponding matrices,
{\it e.g.}, $\gamma^{\alpha\beta} \equiv \gamma^{[\alpha}\gamma^{\beta]}$,
and the gamma matrices with curved indices are defined by multiplying
with the Vierbein, {\it e.g.}, $\gamma^{\mu} \equiv e^{\mu}_{\alpha}
\gamma^{\alpha}$.  Correspondingly, the spinor index $A$ of an 11-d
spinor $\varepsilon^A$ is decomposed into $A = ({\bf a}, {\bf m})$,
{\it i.e.}, $\varepsilon^A = \varepsilon^{({\bf a}, {\bf m})}$,
where ${\bf a}=1,...,4$ is the spinor index for a four component 4-d
spinor and ${\bf m}=1,...,8$ is the index for the spinor representation
of the group $SO(7)$.

With a choice of spherical Ans\" atze given by (\ref{met4d}) through
(\ref{elec}), the Killing spinor equations, which are obtained by setting
the gravitino supersymmetry transformation (\ref{sstran4d}) equal to
zero, are of the following form:
\begin{equation}
{1\over 4}\lambda^{\prime} \gamma^{03} \varepsilon
- {1\over 4}\lambda \partial_r \varphi \gamma^{03} \varepsilon
+{i \over {12}} {{e^i_a e^j_b P_{ij}} \over {Re^{-{\varphi \over 2}}}}
\lambda^{1\over 2} \gamma^{120} \gamma^{ab} \varepsilon
-{i \over 6} {{e^a_i e^b_j Q^{ij}} \over {Re^{\varphi \over 2}}}
\lambda^{1\over 2}\gamma^3 \gamma^{ab} \varepsilon = 0,
\label{tkilling}
\end{equation}
\begin{equation}
\partial_\theta \varepsilon - {1\over 4} \lambda^{1\over 2}{R^{\prime}
\over R^{1/2}} \gamma^{13} \varepsilon + {1\over 4} \sqrt{\lambda R}
\partial_r \varphi \gamma^{13} \varphi -
{i \over {12}}{{e^a_i e^b_j Q^{ij}} \over {Re^{\varphi \over 2}}}
R^{1\over 2} \gamma^{031} \gamma^{ab} \varepsilon -
{i\over 6} {{e^i_a e^j_b P_{ij}} \over {Re^{-{\varphi \over 2}}}}
R^{1\over 2}\gamma^2 \gamma^{ab}\varepsilon = 0,
\label{thkilling}
\end{equation}
\begin{eqnarray}
\partial_{\phi} \varepsilon &-& {1\over 2}{\rm cos}\theta \gamma^{21}
\varepsilon - {1\over 4}\lambda^{1\over 2}{R^{\prime} \over R^{1/2}}
{\rm sin}\theta \gamma^{23} \varepsilon +
{1\over 4} \sqrt{\lambda R}{\rm sin}\theta \partial_r \varphi
\gamma^{23} \varepsilon - {i \over {12}}{{e^a_i e^b_j Q^{ij}} \over
{Re^{\varphi \over 2}}}R^{1\over 2}{\rm sin}\theta \gamma^{032}
\gamma^{ab}\varepsilon
\nonumber \\
&+& {i \over 6}{{e^i_a e^j_b P_{ij}} \over {Re^{-{\varphi \over s}}}}
R^{1\over 2}{\rm sin}\theta \gamma^1 \gamma^{ab} \varepsilon = 0,
\label{phkilling}
\end{eqnarray}
\begin{eqnarray}
\partial_r \varepsilon &+& {1\over 8}(e^l_b \partial_r e_{lc} -
e^l_c \partial_r e_{lb})\gamma^{bc} \varepsilon
-{i \over {12}}{{e^i_a e^j_b P_{ij}} \over {Re^{-{\varphi \over 2}}}}
\lambda^{-{1\over 2}}\gamma^{123} \gamma^{ab}\varepsilon
+ {i \over 6}{{e^a_i e^b_j Q^{ij}} \over {Re^{\varphi \over 2}}}
\lambda^{-{1\over 2}} \gamma^0 \gamma^{ab}\varepsilon = 0,
\label{rkilling}
\end{eqnarray}
\begin{eqnarray}
-{1\over 4}(e^k_d &\partial_r& e_{kb} + e^k_b \partial_r e_{kd})
\lambda^{1\over 2}\gamma^{35} \gamma^b \varepsilon +
{i \over {12}}{{e^a_i e^b_j Q^{ij}} \over {Re^{\varphi \over 2}}}
\gamma^{035} \gamma^{ab}_{\ \ d}\varepsilon +
{i \over {12}} {{e^i_a e^j_b P_{ij}} \over {Re^{-{\varphi \over 2}}}}
\gamma^{125} \gamma^{ab}_{\ \ d}\varepsilon
\nonumber \\
&-& {i \over 3}{{e^d_k e^b_l Q^{kl}} \over {Re^{\varphi \over 2}}}
\gamma^{035} \gamma^b \varepsilon -
{i \over 3}{{e^k_d e^l_b P_{kl}} \over {Re^{-{\varphi \over 2}}}}
\gamma^{125} \gamma^b \varepsilon = 0,
\label{dkilling}
\end{eqnarray}
where the last equation (\ref{dkilling}) is obtained from $e^k_d \delta
\psi_k = 0$ and the prime denotes the differentiation with respect to $r$.

\subsubsection{Constraints on Charges}

Before attempting to solve the Killing spinor equations, we would like to
derive the constraints on charges $P_{ij}$ and $Q^{ij}$ for a general
supersymmetric configuration.

First, one has to determine, by now, a standard form of the angular
coordinate $(\theta, \phi)$ dependence of the spinors $\varepsilon^{\bf m}$
for static, spherically symmetric configurations.  One multiplies
(\ref{thkilling}) by $\gamma^1 {\rm sin}\theta$ and $(\ref{phkilling})$
by $\gamma^2$, subtracts the two resultant equations, and multiplies
the result by $\gamma^2$, thus yielding the following equation:
\begin{equation}
[2\partial_{\phi} + \gamma^1 \gamma^2 {\rm cos}\theta -
2(\gamma^1 \gamma^2 {\rm sin} \theta) \partial_\theta]
\varepsilon^{\bf m} = 0,
\label{sphdep}
\end{equation}
which fixes the angular coordinate dependence of the spinors to be
\begin{equation}
(\varepsilon^{1, {\bf m}}_{u, \ell}, \varepsilon^{2, {\bf m}}_{u,\ell}) =
e^{i\sigma^2 \theta /2} e^{i\sigma^3 \phi /2}(a^{1, {\bf m}}_{u,\ell}(r),
a^{2, {\bf m}}_{u,\ell}(r)) ,
\label{angdep}
\end{equation}
where $\varepsilon^{\bf m}_{u, \ell}$ are the upper (or lower) two
components of the four component spinor $\varepsilon^{\bf m}$,
{\it i.e.}, $(\varepsilon^{\bf m})^T = (\varepsilon^{\bf m}_u ,
\varepsilon^{\bf n}_\ell )$, and $a^{\bf m}_{u,\ell}(r)$ are the
corresponding two component spinors that depend on the radial
coordinate $r$ only.

Then, the Killing spinor equations (\ref{tkilling}), (\ref{thkilling}) and
(\ref{dkilling}), supplemented by (\ref{angdep}), assume the following form:
\begin{equation}
{1 \over \sqrt{\lambda}}[4\sqrt{\lambda R} - 2\lambda \partial_r R -
R\partial_r \lambda + 3\lambda R \partial_r \varphi]\varepsilon_{u,\ell} =
\pm {{\bf P}}_{ab}\gamma^{ab} \varepsilon_{\ell, u} ,
\label{teqn}
\end{equation}
\begin{equation}
{1\over \sqrt{\lambda}}[2\sqrt{\lambda R} - \lambda\partial_r R -
2R\partial_r \lambda + 3\lambda R \partial_r \varphi]\varepsilon_{u,\ell} =
i{{\bf Q}}_{ab}\gamma^{ab}\varepsilon_{\ell,u} ,
\label{theqn}
\end{equation}
\begin{equation}
{1\over 4}{\cal A}_{db}\gamma^b \varepsilon_{\ell, u} \pm {1\over {12}}
({ {\bf P}}_{ab} \mp i {\bf Q}_{ab})\gamma^{ab}_{\ \ d}
\varepsilon_{u,\ell} \mp {1\over 3}({\bf P}_{db} \mp
i{\bf Q}_{db})\gamma^b \varepsilon_{u,\ell} = 0 ,
\label{deqn}
\end{equation}
where ${\bf P}_{ab} \equiv e^{\varphi \over 2} e^i_a e^j_b P_{ij}$,
${\bf Q}_{ab} \equiv e^{-{\varphi \over 2}} e^a_i e^b_j Q^{ij}$,
and ${\cal A}_{ab} \equiv R\lambda^{1\over 2}(e^k_a \partial_r e_{kb} +
e^k_b \partial_r e_{ka})$.  Here, the upper [lower] signs of the
equations are associated with the first [second] subscripts of the
two component spinors.

We are now able to derive constraints on charges $P_{ij}$ and $Q^{ij}$.
These constraints are obtained by ensuring that the gravitino
Killing spinor equations (\ref{teqn}) and (\ref{theqn}), which do not
explicitly depend on the radial derivatives of the Siebenbein, are
satisfied.  After a suitable manipulation
\footnote{The manipulation is similar to the one of Ref. \cite{ADD}.}
of (\ref{teqn}) and (\ref{theqn}), and using the corresponding
anti-commutation relations of $\gamma^{ab}$ matrices, {\it i.e.},
$\{\gamma^{ab},\gamma^{cd}\} = 2\gamma^{abcd}-2(\eta^{ad}\eta^{bc}-
\eta^{ac}\eta^{bd})$, one arrives at the following set of the first
order differential equations
\begin{eqnarray}
{1\over \sqrt{\lambda}}[4\sqrt{\lambda R} - 2\lambda\partial_r R -
R\partial_r \lambda + 3\lambda R \partial_r \varphi] &=&
2\eta_P[\sum_{a<b}({\bf P}_{ab})^2]^{1/2},
\nonumber \\
{1\over \sqrt{\lambda}}[2\sqrt{\lambda R} - \lambda \partial_r R -
2R\partial_r \lambda + 3\lambda R \partial_r \varphi] &=&
2\eta_Q [\sum_{a<b}({\bf Q}_{ab})^2]^{1/2} \ \ \
(\eta_{P,Q} \equiv \pm 1),
\label{firsteqns}
\end{eqnarray}
and the following constraints on the charge configuration:
\begin{equation}
\sum {\bf P}_{ab} {\bf P}_{cd} \gamma^{abcd} = 0 =
\sum {\bf Q}_{ab} {\bf Q}_{cd} \gamma^{abcd},\ \ \ \ \
\sum_{i \neq j} { P}_{ij} {Q}^{ij} = 0 =
\sum {\bf P}_{ab} {\bf Q}_{cd} \gamma^{abcd} \ \ ,
\label{chcon}
\end{equation}
as well as the constraint between the upper and lower two-component spinors:
\begin{equation}
\varepsilon_u=\eta_P [4\sum_{a'<b'}({\bf P}_{a'b'})^2]^{-1/2}
{\bf P}_{ab} \gamma^{ab}\varepsilon_{\ell}\ \ (\eta_{P} \equiv \pm 1).
\end{equation}

We can find a constraint on charges by analyzing the constraints
(\ref{chcon}) on the asymptotic values ${\bf P}_{ab\,\infty}$ and
${\bf Q}_{ab\,\infty}$ of ${\bf P}_{ab}$ and ${\bf Q}_{ab}$.  Manifest
$SL(7,\Re)$ symmetry, {\it i.e.}, the rescaling symmetry and the $SO(7)$
rotations, allows one to bring the asymptotic value of the internal metric
to the form $g_{ij\,\infty}= -\delta_{ij}$, without loss of generality.
Thus, the asymptotic value of the Siebenbein can be chosen to be
$e^i_{a\,\infty}\equiv \delta^i_a$, up to global $SO(7)$ rotations and
the rescaling of the radii of the internal coordinates.  In the case of
$e^i_{a\,\infty}\equiv \delta^i_a$, the only charge configurations, which
satisfy the constraints (\ref{chcon}), are of the following two types:
\begin{itemize}
\item
The only non-zero charges are $P_{ij}\ne 0$ and $Q^{ij}\ne 0$,
subject to the constraint $\sum_j P_{ij}Q^{ij}=0$, where $i$ corresponds
to a fixed choice of the index.
\item
The only allowed nonzero charges are $(P_{ij}, P_{ik}, P_{jk})\ne 0$
and $(Q^{ij}, Q^{ik}, Q^{jk})\ne 0$, subject to the constraint
$P_{ij}Q^{ij}+P_{ik}Q^{ik}+P_{jk}Q^{jk}=0$, where  $i\ne j\ne k$
correspond to a fixed choice of three indices.
\end{itemize}

The first charge configuration can be transformed, by a subset of
$SO(6)\subset SO(7)$ rotations, into a form in which the only nonzero
charges are $P_{ij}\ne 0$ and  $Q^{ik}\ne 0$, where $i\ne j\ne k$ correspond
to a fixed choice of the three indices.  The second charge configuration
can also be transformed, by a subset of $SO(3)\subset SO(7)$ rotations,
in the same form with only nonzero charges given by $P_{ij}\ne 0$ and
$Q^{ik}\ne 0$.

A generating  solution is, therefore, the one with one electric charge,
say  $Q^{ij}$, and one magnetic charge, say $P_{ik}$.  Thus, the
most general supersymmetric configuration can be obtained by imposing a
subset of global $SO(7)$ transformations, {\it i.e.}, $SO(7)/SO(3)$
transformations with $3n-6=15$ parameters, on a generating solution with
one electric charge, say  $Q^{ij}$, and one magnetic charge, say $P_{ik}$
(2 parameters).  Consequently, the most general supersymmetric charge
configuration is of {\it constrained} one; among non-zero $n(n-1)/2=21$
electric charges and $n(n-1)/2=21$ magnetic charges, only $(3n-6)+2=17$ are
independent.

\subsubsection{Supersymmetric Three-Form Black Hole Solutions with Diagonal
Internal Metric}

Our next goal is to obtain the explicit form of the generating solution with
only one magnetic charge $P_{ij}$ and one electric charge $Q^{ik}$ non-zero.
This type of solution will be obtained with a diagonal internal metric
Ansatz, {\it i.e.}, the Siebenbein is chosen to be of the form:
\begin{equation}
e^k_a = {\rm diag}(e_1,...,e_7).
\end{equation}

In this case, the first order differential equations (\ref{teqn}) $-$
(\ref{deqn}) reduce to the following simplified form:
\begin{equation}
{1\over \sqrt{\lambda}}[4\sqrt{\lambda R} - 2\lambda\partial_r R -
R\partial_r \lambda +3\lambda R \partial_r \varphi] =
2\eta_P {\bf P}_{\hat i\hat j},
\label{teq}
\end{equation}
\begin{equation}
{1\over \sqrt{\lambda}}[2\sqrt{\lambda R} - \lambda \partial_r R -
2R\partial_r \lambda + 3\lambda R \partial_r \varphi] =
2\eta_Q {\bf Q}_{\hat i\hat k},
\label{theq}
\end{equation}
\begin{equation}
-{1\over 4}{\cal  A}_{\hat{i}\hat{i}} \varepsilon_{\ell,u} \mp {1\over 3}
{\bf P}_{\hat{i}\hat{j}}\Gamma^{\hat{i}\hat{j}}\varepsilon_{u,\ell} +
{i\over 3}{\bf Q}_{\hat{i}\hat{k}} \Gamma^{\hat{i}\hat{k}}
\varepsilon_{u,\ell} = 0,
\label{ieq}
\end{equation}
\begin{equation}
-{1\over 4}{\cal A}_{\hat{j}\hat{j}} \varepsilon_{\ell,u} \mp {1\over 3}
{\bf P}_{\hat{i}\hat{j}}\Gamma^{\hat{i}\hat{j}}\varepsilon_{u,\ell} -
{i\over 6}{\bf Q}_{\hat{i}\hat{k}} \Gamma^{\hat{i}\hat{k}}
\varepsilon_{u,\ell} = 0,
\label{jeq}
\end{equation}
\begin{equation}
-{1\over 4}{\cal A}_{\hat{k}\hat{k}} \varepsilon_{\ell,u} \pm {1\over 6}
{\bf P}_{\hat{i}\hat{j}}\Gamma^{\hat{i}\hat{j}}\varepsilon_{u,\ell} +
{i\over 3}{\bf Q}_{\hat{i}\hat{k}} \Gamma^{\hat{i}\hat{k}}
\varepsilon_{u,\ell} = 0,
\label{keq}
\end{equation}
\begin{equation}
-{1\over 4}{\cal A}_{\hat{\ell}\hat{\ell}} \varepsilon_{\ell,u} \pm {1\over 6}
{\bf P}_{\hat{i}\hat{j}}\Gamma^{\hat{i}\hat{j}}\varepsilon_{u,\ell} -
{i\over 6}{\bf Q}_{\hat{i}\hat{k}} \Gamma^{\hat{i}\hat{k}}
\varepsilon_{u,\ell} = 0 \ \ \ \ \ (\hat{\ell} \neq \hat{i},
\hat{j}, \hat{k}),
\label{leq}
\end{equation}
where ${\cal A}_{\hat{n} \hat{n}} \equiv 2R\lambda^{1\over 2}\partial_r
{\rm ln} e_{\hat{n}}$ and the hats on the curved indices denote the
corresponding flat indices.

We shall now solve the equations (\ref{teq}) $-$ (\ref{leq}) to get the
explicit dyonic supersymmetric solutions with one magnetic $P_{ij}$ and
one electric $Q_{ik}$ charges
\footnote{The derivation for supersymmetric as well as non-extreme solutions
are along the similar lines as those for $U(1)_M\times U(1)_E$ supersymmetric
\cite{SUPER} and non-extreme\cite{NONEX,PARK,ALLKKBH} KK BH's.}.
First, from (\ref{ieq}) $-$ (\ref{leq}), we obtain the following relations
among Siebenbein components:
\begin{equation}
\check{e}_{\hat{i}} = (\check{e}_{\hat{\ell}})^{-2}, \ \ \ \ \ \
\check{e}_{\hat{j}} \check{e}_{\hat{k}} \check{e}_{\hat{\ell}} = 1,
\label{screl1}
\end{equation}
where $\check{e}_{\hat i} \equiv e_{\hat i}/e_{\hat{i}\,\infty}$.
Substituting the difference between (\ref{ieq}) and (\ref{jeq}) into
(\ref{theq}), we obtain
\begin{equation}
\check{e}_{\hat{i}}^2 \check{e}_{\hat{j}}^{-2} =
e^{3(\varphi-\varphi_\infty)}\lambda^{-1}.
\label{screl2}
\end{equation}
Making use of (\ref{screl1}), we arrive at the relation
\begin{equation}
e^{(\varphi-\varphi_\infty)} = {\rm det}\,\check{e}^d_m = \check{e}^
{-1}_{\hat{\ell}} \ \ \ (\hat{\ell} \neq \hat{i}, \hat{j}, \hat{k}).
\label{screl3}
\end{equation}
Then, the following relation between the 4-d metric components $\lambda(r)$
and $R(r)$ can be obtained  by substituting (\ref{teq}) and (\ref{theq})
into (\ref{leq}), making use of (\ref{screl3}):
\begin{equation}
2\sqrt{\lambda R} = \lambda\partial_r R + R\partial_r \lambda,
\label{meteq}
\end{equation}
which can be solved to yield
\begin{equation}
\lambda R = (r - r_H)^2,
\label{metrel}
\end{equation}
where the integration constant $r_H$ corresponds to the event horizon,
{\it i.e.}, $\lambda(r_H) = 0$.  This is, again, another standard result
for 4-d, supersymmetric, spherically symmetric, static solutions.

Furthermore, with the help of (\ref{meteq}) we can simplify (\ref{teq})
and (\ref{theq}) to the following forms:
\begin{equation}
\sqrt{\lambda}R({{\partial_r \lambda} \over \lambda} + 3\partial_r
\varphi) = 2\eta_P {\bf P}_{\hat{i}\hat{j}},
\label{eq1}
\end{equation}
\begin{equation}
\sqrt{\lambda}R(-{{\partial_r \lambda} \over \lambda} + 3\partial_r
\varphi) = 2\eta_Q {\bf Q}_{\hat{i}\hat{k}}.
\label{eq2}
\end{equation}
Note the symmetry of the above two equations under the electric-magnetic
duality transformation, {\it i.e.}, ${\bf P}_{\hat{i}\hat{j}}
\leftrightarrow -{\bf Q}_{\hat{i}\hat{k}}$ and $\varphi \to -\varphi$.
By adding these two equations, we obtain the following equation
\begin{equation}
\partial_r \varphi = {1\over {3\sqrt{\lambda}R}}[\eta_P
{\bf P}_{\hat{i}\hat{j}} + \eta_Q {\bf Q}_{\hat{i}\hat{k}}] ,
\label{nohair}
\end{equation}
which is in accordance with a no-hair theorem, {\it i.e.}, when
electromagnetic fields are zero ($P_{ij} = 0 = Q_{ik}$) the dilaton
field $\varphi$ is constant.

The expression relating the 4-d metric component $\lambda$ and the
dilaton $\varphi$ can be obtained by multiplying (\ref{eq1}) by
$\eta_{Q} {\bf Q}_{\hat{i}\hat{k}}$ and (\ref{eq2}) by
$\eta_P {\bf P}_{\hat{i}\hat{j}}$, followed by addition of the
resulting two equations.  The resultant equation can be solved to yield
\begin{equation}
\lambda = {{\eta_P e^{3\varphi}{\bf P}_{ij\,\infty} + \eta_Q
e^{-3\varphi}{\bf Q}_{ik\,\infty}} \over {\eta_P {\bf P}_{ij\,\infty}
+ \eta_Q {\bf Q}_{ik\,\infty}}} ,
\label{metscrel}
\end{equation}
where ${\bf P}_{ij\,\infty} \equiv e^{\varphi_\infty/2}
g^{-{1\over 2}}_{ii\,\infty}g^{-{1\over 2}}_{jj\,\infty}P_{ij}$ and
${\bf Q}_{ik\,\infty} \equiv e^{\varphi_\infty/2}
g^{-{1\over 2}}_{ii\,\infty} g^{-{1\over 2}}_{kk\,\infty}Q_{ik}$ are
the ``screened'' electric and magnetic charges.  Finally, the following
ordinary differential equation for $\varphi$ is obtained by substituting
(\ref{metscrel}) into (\ref{nohair}) and making use of Eqs. (\ref{screl1})
$-$ (\ref{metrel}):
\begin{equation}
\partial_r \varphi = {1 \over {3(r - r_H)}}[\eta_P e^{3\varphi}
{\bf P}_{ij\,\infty} + \eta_Q e^{-{3\varphi}}{\bf Q}_{ik\,\infty}]^2
{1 \over {\eta_P {\bf P}_{ij\,\infty} + \eta_Q {\bf Q}_{ik\,\infty}}} .
\label{dileq}
\end{equation}
Note, again, the symmetry of (\ref{dileq}) under the electric-magnetic
duality transformation.  The explicit solution for the dilaton $\varphi$
is, then, given by
\begin{equation}
e^{3\varphi} = \left ({{r - r_H - 2\eta_Q {\bf Q}_{ik\,\infty}} \over
{r - r_H + 2\eta_P {\bf P}_{ij\,\infty}}} \right )^{1\over 2} =
\left ( {{r - 2|{\bf P}_{ij\,\infty}|} \over {r - 2|{\bf Q}_{ik\,\infty}|}}
\right )^{1\over 2} ,
\label{dilsol}
\end{equation}
where we have identified $r_H = 2\eta_P {\bf P}_{ij\,\infty} - 2\eta_Q
{\bf Q}_{ik\,\infty}$, and chosen the signs of $\eta_{P,Q}$ so that
$\eta_P {\bf P}_{ij\,\infty} = |{\bf P}_{ij\,\infty}|$ and $-\eta_Q
{\bf Q}_{ik\,\infty} = |{\bf Q}_{ik\,\infty}|$.

Now, we are ready to write down the explicit supersymmetric BH solutions.
Making use of the solution (\ref{dilsol}) for the dilaton $\varphi$ and
the relations (\ref{screl1}) $-$ (\ref{screl3}) and (\ref{metscrel})
among various fields,  as well as Eq. (\ref{dkilling}) we obtain the
following result for the scalar fields as well as for the Killing spinor
\footnote{Note, the 3-form BH solution has charges which are different
from those of KK BH's by a factor of 2 because of a different
normalization for the corresponding gauge fields.}:
\begin{eqnarray}
\lambda &=& {{r - 2|{\bf P}_{ij\,\infty}| - 2|{\bf Q}_{ik\,\infty}|} \over
{(r - 2|{\bf P}_{ij\,\infty}|)^{1\over 2}(r - 2|{\bf Q}_{ik\,\infty}|)^
{1\over 2}}} ,
\nonumber \\
R &=& r^2 (1 - {{2|{\bf P}_{ik\,\infty}| +2|{\bf Q}_{ik\,\infty}|}\over r})
(1 - {{2|{\bf P}_{ij\,\infty}|} \over r})^{1 \over 2}
(1 - {{2|{\bf Q}_{ik\,\infty}|} \over r})^{1 \over 2},
\nonumber \\
e^{3(\varphi - \varphi_{\infty})} &=& \left ( {{r - 2|{\bf P}_{ij\,\infty}|}
\over {r - 2|{\bf Q}_{ik\,\infty}|}}\right)^{1\over 2} ,
\nonumber \\
g_{ii}/g_{ii\,\infty} &=& e^{-4(\varphi - \varphi_\infty)} =
\left ( {{r - 2|{\bf Q}_{ik\,\infty}|} \over {r - 2|{\bf P}_{ij\,\infty}|}}
\right)^{2\over 3} ,
\nonumber \\
g_{jj}/g_{jj\,\infty} &=& \lambda^{-1}e^{-(\varphi-\varphi_{\infty})} =
{{(r - 2|{\bf P}_{ij\,\infty}|)^{1\over 3} (r-2|{\bf Q}_{ik\,\infty}|)^
{2\over 3}} \over {r - 2|{\bf P}_{ij\,\infty}| -
2|{\bf Q}_{ik\,\infty}|}} ,
\nonumber \\
g_{kk}/g_{kk\,\infty} &=& \lambda e^{-(\varphi-\varphi_\infty)} =
{{r - 2|{\bf P}_{ij\,\infty}| - 2|{\bf Q}_{ik\,\infty}|} \over
{(r - 2|{\bf P}_{ij\,\infty}|)^{2\over 3} (r-2|{\bf Q}_{ik\,\infty}|)^
{1\over 3}}} ,
\nonumber \\
g_{\ell \ell}/ g_{\ell \ell \,\infty} &=& e^{2(\varphi-\varphi_\infty)} =
\left ( {{r - 2|{\bf P}_{ij\,\infty}|}\over {r - 2|{\bf Q}_{ik\,\infty}|}}
\right)^{1\over 3} \ \ \ (\ell \neq i,j,k),
\nonumber \\
a^{\bf m}_u(r)&=&a^{\bf m}_{u\,\infty}{{(r-2|{\bf P}_{ij\,\infty}|
-2|{\bf Q}_{ik\,\infty}|)^{1\over 4}} \over {(r-2|{\bf P}_
{ij\,\infty}|)^{1\over 6}(r-2|{\bf Q}_{ik\,\infty}|)^{1\over {12}}}},
\label{antisol}
\end{eqnarray}
where, again, ${\bf P}_{ij\,\infty} \equiv e^{\varphi_\infty/2}
g^{-{1\over 2}}_{ii\,\infty}g^{-{1\over 2}}_{jj\,\infty}P_{ij}$ and
${\bf Q}_{ik\,\infty} \equiv e^{\varphi_\infty/2}
g^{-{1\over 2}}_{ii\,\infty} g^{-{1\over 2}}_{kk\,\infty}Q_{ik}$.
The 4-d space-time of these solutions is the same as the one
of supersymmetric $U(1)_M\times U(1)_E$ KK BH's (see Eq. (\ref{kksol})),
provided one makes the replacement $2{\bf P_{ij\,\infty}}\rightarrow
{\bf P}_{j\,\infty}$ and $2{\bf Q}_{ik\,\infty}\rightarrow
{\bf Q}_{k\,\infty}$.  Thus, the corresponding global space-time
and thermal properties are the same.

These supersymmetric solutions saturate the corresponding Bogomol'nyi
bound on the ADM mass.  The derivation of the bound is along the lines
spelled out in Refs.\cite{HARVEY,SUPER}.

\subsubsection{Non-extreme 3-Form Black Hole Solutions with Diagonal
Internal Metric}

We also point out that the corresponding non-extreme solutions can
be obtained by solving explicitly the corresponding second order
differential equations with a diagonal internal metric Ansatz.
Here, we quote the result:
\begin{eqnarray}
\lambda&=&{{(r-r_+)}\over{(r-r_+ +2\hat{\bf P}_{ij\,\infty})^{1\over 2}
(r-r_+ +2\hat{\bf Q}_{ik\,\infty})^{1\over 2}}},
\nonumber \\
R &=& r^2 (1-{{r_+ - 2\beta}\over r})(1-{{r_+ - 2\hat{\bf P}_{ij\,\infty}}
\over r})^{1 \over 2} (1-{{r_+ - 2\hat{\bf Q}_{ik\,\infty}} \over r})^
{1 \over 2},
\nonumber \\
e^{3(\varphi-\varphi_\infty)}& = &\left ( {{r-r_+ +
2\hat{\bf Q}_{ik\,\infty}}\over {r-r_+ +
2\hat{\bf P}_{ij\,\infty}}} \right )^{1\over 2},
\nonumber  \\
g_{ii}/g_{ii\,\infty} &=& \left ( {{r-r_+ + 2\hat{\bf P}_{ij\,\infty}}
\over {r-r_++2\hat{\bf Q}_{ik\,\infty}}} \right )^{2\over 3},
\nonumber  \\
g_{jj}/g_{jj\,\infty}& =& {{(r-r_++2\hat{\bf P}_{ij\,\infty})^{1\over 3}
(r-r_++2\hat{\bf Q}_{ik\,\infty})^{2\over 3}} \over {(r-r_+)}},
\nonumber \\
g_{kk}/g_{kk\,\infty} &=& {{(r-r_+)}\over {(r-r_++2\hat{\bf P}_
{ij\,\infty})^{2\over 3}(r-r_++2\hat{\bf Q}_
{ik\,\infty})^{1\over 3}}}, \nonumber \\
 g_{\ell\ell}/g_{\ell\ell\,\infty}& =&
\left ( {{r-r_++2\hat{\bf Q}_{ik\,\infty}}\over {r-r_++
2\hat{\bf P}_{ij\,\infty}}} \right )^{1\over 3}
(\ell \neq i,j,k),
\label{nonext}
\end{eqnarray}
where $({\bf P}_{ij\,\infty})^2 = \hat{\bf P}_{ij\,\infty}
(\hat{\bf P}_{ij\,\infty} - \beta)$, $({\bf Q}_{ik\,\infty})^2 =
\hat{\bf Q}_{ik\,\infty}(\hat{\bf Q}_{ik\,\infty}-\beta)$,
$r_+=\beta+(|{\bf P}_{ij\,\infty}|\sqrt{\beta^2+4{\bf P}^2_
{ij\,\infty}} -|{\bf Q}_{ik\,\infty}|\sqrt{\beta^2+4{\bf Q}_{ik\,\infty}^2})/
(|{\bf P}_{ij\,\infty}|-|{\bf Q}_{ik\,\infty}|)$, and  $\beta > 0$
is the non-extremality parameter.  The ADM mass $M$ of the configuration,
the Hawking temperature $T_H$, and the entropy $S$ are given by:
\begin{eqnarray}
M&=&2\beta + \sqrt{4{\bf P}^2_{ij\,\infty} +
\beta^2} + \sqrt{4{\bf Q}^2_{ik\,\infty} + \beta^2},
\\
T_H&=&1/(4\pi[\beta+(4{\bf P}^2_{ij\,\infty}+
\beta^2)^{1\over 2}]^{1\over 2}[\beta+(4{\bf Q}^2_{ik\,\infty}+
\beta^2)^{1\over 2}]^{1\over 2}),
\\
S&=&2\pi\beta[\beta+(4{\bf P}_{ij\,\infty}^2+\beta^2)^{1\over 2}]^
{1\over 2}[\beta+(4{\bf Q}_{ik\,\infty}^2+\beta^2)^{1\over 2}]^{1\over 2}.
\label{propenex}
\end{eqnarray}
The global space-time is that of Schwarzschild BH's, with a horizon at
$r=r_+$ and a space-like singularity at $r=r_+-2\beta$.  The 4-d
space-time of the above solution is the same as the one of non-extreme
$U(1)_M\times U(1)_E$ KK BH's \cite{NONEX}, with the replacement
$2{\bf P}_{ij\,\infty}\rightarrow {\bf P}_{j\,\infty}$ and
$2{\bf Q}_{ik\,\infty}\rightarrow {\bf Q}_{k\,\infty}$.

\subsubsection{Symmetry Tranformations between Kaluza-Klein and
Three-Form Black Hole Solutions}

Solutions for supersymmetric (see Eq. (\ref{antisol})) and non-extreme
(see Eq. (\ref{nonext})) 3-form BH's with a diagonal internal metric
bear similarities with the corresponding solutions for supersymmetric
(see Eq. (\ref{kksol})) and non-extreme \cite{NONEX} KK BH's.
In particular, the 4-d metric $g_{\mu\nu}$ and thus the global space-time
as well as thermal properties of both types of BH's are the same.  This is
an indication that the two types of the solutions are related by a
discrete subset of $E_7$ transformation.  A discrete symmetry, which
transforms the effective action of $U(1)_M\times U(1)_E$ KK BH's into
the corresponding one for the 3-form BH's, relates the charges
$P_{ij}, Q^{ik}$ and the normalized (diagonal) metric coefficients
${\check{g}^{\prime}_{ii}}\equiv g^{\prime}_{ii}/g^{\prime}_{ii\,\infty}$
of the 3-form effective action to charges  $P_{j}, Q^{k}$ and the normalized
(diagonal) metric coefficients ${\check{g}_{ii}}\equiv g_{ii}/g_{ii\,\infty}$
of the KK effective action in the following way:
\begin{eqnarray}
P_{ij}&=&{P_j \over 2},\ \ \ \ \ \  Q^{ik}={Q^k \over 2},
\nonumber\\
{{\check{g}^{\prime}_{jj}}\over {\check{g}^{\prime}_{kk}}} =
{\check{g}_{kk} \over \check{g}_{jj}},\ \
\check{g}^{\prime}_{jj}\check{g}^{\prime}_{kk}=(\check{g}_{kk}
&\check{g}_{jj}&)^{-1/3}\check{g}^{\prime}_{ii}= (\check{g}_{jj}
\check{g}_{kk})^{-2/3},\ \  \prod_{\ell\neq (ijk)}
\check{g}^{\prime}_{\ell\ell}=(\check{g}_{jj}\check{g}_{kk})^{4/3}.
\label{trans}
\end{eqnarray}

The above discrete transformation in turn corresponds to a subset of
continuous transformations, which preserve the ``length'' of a
combined KK and 3-form ``charge vector'', and which involve the
psueudoscalar field as well
\footnote{A subset of such transformations relates the fields
within NS-NS sector of type-IIA string compactified on a 6-torus (see
the subsequent chapter), which is the same as a subsector of
heterotic string (associated with 6 $U(1)$ KK gauge fields and
6 $U(1)$ gauge fields originating from the 10-d two-form field).
In this subsector, the fields are related by the continuous $SO(6,6)$
transformations \cite{SEN1} which also involve 6 pseudo-scalar fields.}.

\section{Type-IIA superstring theory and eleven-dimensional supergravity}

In this chapter, we relate the 4-d supersymmetric BH solutions of 11-d SG,
which were discussed in the previous chapter, to the corresponding BH's of
type-IIA superstring theory compactified on a 6 torus.  In particular, we
would like to relate the scalar field degrees of 11-d SG, parameterizing the
internal space-time, to the string coupling of the type-IIA superstring
compactified on a 6-torus, and thus, by virtue of duality between the two
theories, obtain information on mass spectrum of strongly coupled type-IIA
string compactified on a 6-torus.

The zero slope limit of the type-IIA 10-d superstring theory can be
obtained by dimensional reduction of 11-d SG on a circle $S^1$\cite{HUQ}.
This can be accomplished by choosing the following triangular gauge form
for the Elfbein $E^{(11)\, A}_M$:
\begin{equation}
E^{(11)\, A}_M = \left ( \matrix{e^{-{\Phi \over 3}}
e^{(10)\, \breve{\alpha}}_{\breve{\mu}} & e^{{2\over 3}\Phi}
B_{\breve{\mu}} \cr 0 & e^{{2\over 3}\Phi}} \right ) ,
\label{kktoten}
\end{equation}
where $\Phi$ corresponds to the 10-d dilaton field in NS-NS sector of
the superstring theory, $e^{(10)\, \breve{\alpha}}_{\breve{\mu}}$ is the
Zehnbein in NS-NS sector, and $B_{\breve{\mu}}$ corresponds to a one-form
in RR sector.  Here, the breve denotes the 10-d space-time vector index.
And the 3-form $A^{(11)}_{MNP}$ is decomposed into $A^{(11)}_{MNP} =
(A_{\breve{\mu}\breve{\nu}\breve{\rho}}, A_{\breve{\mu}\breve{\nu} 11}
\equiv A_{\breve{\mu}\breve{\nu}})$, where $A_{\breve{\mu}\breve{\nu}
\breve{\rho}}$ is identified as a 3-form in RR sector and
$A_{\breve{\mu}\breve{\nu}}$ is the antisymmetric tensor in NS-NS sector.
Then, 11-d bosonic action (\ref{action11d}) becomes the following
10-d, $N=2$ SG action:
\begin{equation}
{\cal L} = {\cal L}_{NS} + {\cal L}_R ,
\label{sg10d}
\end{equation}
with
\begin{eqnarray}
{\cal L}_{NS} &=& -{1\over 4}e^{(10)}e^{-2\Phi}[{\cal R} +
4\partial_{\breve{\mu}}\Phi \partial^{\breve{\mu}}\Phi -
{1\over 3}F_{\breve{\mu}\breve{\nu}\breve{\rho}}
F^{\breve{\mu}\breve{\nu}\breve{\rho}}],
\nonumber \\
{\cal L}_R &=& -{1 \over 4}e^{(10)}[{1\over 4}G_{\breve{\mu}\breve{\nu}}
G^{\breve{\mu}\breve{\nu}} +{1\over {12}}F^{\prime}_{\breve{\mu}\breve{\nu}
\breve{\rho}\breve{\sigma}} F^{\prime\, \breve{\mu}\breve{\nu}\breve{\rho}
\breve{\sigma}} - {6 \over {(12)^3}}\varepsilon^{\breve{\mu}_1 \cdots
\breve{\mu}_{10}} F_{\breve{\mu}_1 \cdots \breve{\mu}_4}
F_{\breve{\mu}_5 \cdots \breve{\mu}_8} A_{\breve{\mu}_9 \breve{\mu}_{10}}],
\label{nsr}
\end{eqnarray}
where $e^{(10)} \equiv {\rm det}\, e^{(10)\, \breve{\alpha}}_{\breve{\mu}}$,
$\cal R$ is the Ricci scalar defined in terms of the Zehnbein,
$F_{\breve{\mu}\breve{\nu}\breve{\rho}} \equiv 3\partial_{[\breve{\mu}}
A_{\breve{\nu}\breve{\rho}]}$, $G_{\breve{\mu}\breve{\nu}} \equiv
2\partial_{[\breve{\mu}}B_{\breve{\nu}]}$, $F^{\prime}_{\breve{\mu}\breve{\nu}
\breve{\rho}\breve{\sigma}} \equiv 4\partial_{[\breve{\mu}}A_{\breve{\nu}
\breve{\rho}\breve{\sigma}]} - 4F_{[\breve{\mu}\breve{\nu}\breve{\rho}}
B_{\breve{\sigma}]}$, and $\varepsilon^{\breve{\mu}_1 \cdots \breve{\mu}_{10}}
\equiv \varepsilon^{\breve{\mu}_1 \cdots \breve{\mu}_{10} 11}$.
The ferminoic sector in 10-d contains Majorana gravitino $\psi_{\breve{\mu}}$
and fermion $\psi_{11}$ that come from the 11-d gravitino $\psi^{(11)}_M$,
{\it i.e.}, $\psi^{(11)}_M = (\psi_{\breve{\mu}}, \psi_{11})$.  These spinors
can be split into two Majorana-Weyl spinors of left- and right-helicities.

In order to obtain the effective 4-d action of the type-IIA superstring
compactified on a 6-torus, one chooses the following KK Ansatz for the
Zehnbein:
\begin{equation}
e^{(10)\, \breve{\alpha}}_{\breve{\mu}} =
\left ( \matrix{e^{\alpha}_{\mu} & \bar{B}^m_{\mu}\bar{e}^a_m \cr
0 & \bar{e}^a_m} \right ) ,
\label{zehnbein}
\end{equation}
where $\bar{B}^m_{\mu}$ ($m=1,...,6$) are Abelian KK gauge fields,
$e^{\alpha}_{\mu}$ is the string frame 4-d Vierbein and $\bar{e}^a_m$
is the Sechsbein.   In the following, we shall set all the other scalars,
except those associated with the Sechsbein $\bar{e}^a_m$ and the 10-d
dilaton $\Phi$, to zero
\footnote{We turn off the scalar fields $B_{m}$ ($m$=4,...,9) associated
with the 10-d $U(1)$ gauge field $B_{\breve{\mu}}$.  These fields are
related to the internal metric coefficients $g_{m7}$ ($m=1,\cdots ,6)$
of 11-d SG.}.
In this case, the string-frame 4-d bosonic action for the type-IIA
superstring is of the following form:
\begin{eqnarray}
{\cal L}_{II} = &-&{1\over 4}e[e^{-2\phi}({\cal R} + 4\partial_{\mu}
\phi \partial^{\mu}\phi + {1\over 4}\partial_{\mu}\bar{g}_{mn}
\partial^{\mu}\bar{g}^{mn} - {1\over 4}\bar{g}_{mn}\bar{G}^m_{\mu\nu}
\bar{G}^{n\, \mu\nu} -\bar{g}^{mn}{\bar F}_{\mu\nu\, m}
{\bar F}^{\mu\nu}_{\ \ n})
\nonumber \\
&+&{1\over 4}e^{{\bar \sigma}}\bar{G}_{\mu\nu}\bar{G}^{\mu\nu} +
{1\over 2}e^{{\bar \sigma}} \bar{g}^{mn}\bar{g}^{pq}
\bar{F}_{\mu\nu\, mp}\bar{F}^{\mu\nu}_{\ \ nq}] ,
\label{type2}
\end{eqnarray}
where $e \equiv {\rm det}\, e^{\alpha}_{\mu}$, $2\phi \equiv 2\Phi -
{\rm ln}\, {\rm det}\, \bar{e}^a_m$ (parameterizing the {\it string
coupling}), ${\bar \sigma} \equiv {\rm ln}\, {\rm det}\, \bar{e}^a_m$
(parameterizing the volume of 6-torus), $\bar{g}_{mn} \equiv \eta_{ab}
\bar{e}^a_m \bar{e}^b_n = -\bar{e}^a_m \bar{e}^a_n$, and $\bar{G}^m_{\mu\nu}
\equiv \partial_{\mu} \bar{B}^m_{\nu} - \partial_{\nu} \bar{B}^m_{\mu}$.
Here, the field strengths $\bar{F}_{\mu\nu\, m}$, $\bar{G}_{\mu\nu}$ and
$\bar{F}_{\mu\nu\, mn}$ are defined in terms of the Abelian gauge fields
decomposed from 10-d two-form $A_{\breve{\mu}\breve{\nu}}$, one-form
$B_{\breve{\mu}}$ and the three-form $A_{\breve{\mu}\breve{\nu}
\breve{\rho}}$ fields, respectively.  The following Einstein-frame
action can be obtained by the Weyl rescaling $g_{\mu\nu} \to g^E_{\mu\nu}
= e^{-2\phi}g_{\mu\nu}$:
\begin{eqnarray}
{\cal L}_{II} = &-&{1\over 4}e^E[{\cal R}^E - 2\partial_{\mu}\phi
\partial^{\mu} \phi + {1\over 4}\partial_{\mu}\bar{g}_{mn}
\partial^{\mu}\bar{g}^{mn} - {1\over 4}e^{-2\phi}\bar{g}_{mn}
\bar{G}^m_{\mu\nu}\bar{G}^{n\,\mu\nu}-e^{-2\phi}\bar{g}^{mn}
{\bar F}_{\mu\nu\, m}{\bar F}^{\mu\nu}_{\ \ n}
\nonumber \\
&+& {1\over 4}e^{\bar \sigma}\bar{G}_{\mu\nu}\bar{G}^{\mu\nu}
+ {1\over 2}e^{\bar \sigma}\bar{g}^{mn}\bar{g}^{pq}
\bar{F}_{\mu\nu\, mp}\bar{F}^{\mu\nu}_{\ \ nq}],
\label{eintype2}
\end{eqnarray}
where $e^E \equiv \sqrt{-{\rm det}\,g^E_{\mu\nu}}$ and ${\cal R}^E$
is the Ricci scalar defined in terms of the Einstein-frame metric
$g^E_{\mu\nu}$.

Since we have turned off the scalar fields associated with the 10-d
$U(1)$ gauge field $B_{\breve{\mu}}$, thus the internal metric
coefficients $g_{m7}$ of 11-d SG, the $SO(7)$ symmetry among 7 KK
gauge fields and among 21 3-form  gauge fields, separately, breaks down
to the $SO(6)$ symmetry, which {\it do not mix} the gauge fields of RR
and NS-NS sectors.  The RR sector consists of one KK gauge field
$\bar{B}_\mu$, which transforms as a singlet of $SO(6)$, and fifteen
3-form $U(1)$ gauge fields $\bar{A}_{\mu\, mn}$, which transform as
{\bf 15} antisymmetric representation of $SO(6)$. The NS-NS sector
consists of six KK gauge fields $\bar{B}^m_\mu$ and six 3-form $U(1)$
gauge fields $\bar{A}_{\mu\, n}$, each of them transforming as a
${\bf 6}$ vector representation of $SO(6)$.
\footnote{In order to have the full manifestation of the $SO(7)$
symmetry of 11-d SG in the BH solutions of type-IIA superstring,
the scalar fields which are associated with the 10-d $U(1)$ gauge field
$B_{\breve{\mu}}$ has to be included.}.

Given two classes (\ref{kksol}) and (\ref{antisol}) of supersymmetric
solutions in 11-d SG, one can find the corresponding solutions in the
type-IIA superstring which are associated with different Abelian gauge
fields in (\ref{type2}).  For this purpose, one has to relate the bosonic
fields in 4-d type-IIA superstring action (\ref{eintype2}) to the ones
in 4-d action (\ref{action4d}) of compactified 11-d SG.  This is done by
keeping track of field decomposition and redefinition, and by comparing
the compactification Ans\" atze in two different schemes, {\it i.e.},
one corresponding to 11-d $\to$ 10-d $\to$ 4-d and the other one
corresponding to 11-d $\to$ 4-d.  The expressions of fields in the 4-d
type-IIA superstring action (\ref{eintype2}) in terms of those in the
4-d action (\ref{action4d}) of 11-d SG are given by
\begin{eqnarray}
\phi &=& -{3\over 7}{\varphi} + {1 \over 4}{\rm ln}\, \rho_{77},\ \ \
\bar{\sigma} = {9\over 7}\varphi + {\rm ln} \, \rho_{77} ,\ \ \
\bar{\rho}_{mn} = (\rho_{77})^{1\over 6}\rho_{mn},
\nonumber \\
\bar{B}^m_{\mu} &=& B^m_{\mu}, \ \ \ \  \bar{B}_{\mu} = B^7_{\mu} , \ \ \
\bar{A}_{\mu m} = A_{\mu m 7}, \ \ \ \  \bar{A}_{\mu\,mn} = A_{\mu\,mn},
\label{rel}
\end{eqnarray}
where $m,n = 1,...,6$ and $\bar{\rho}_{mn}$ is the unimodular part of the
internal metric $\bar{g}_{mn}$ ($\bar{g}_{mn} = -e^{\bar{\sigma}/{3}}
\bar{\rho}_{mn}$).  Recall, the internal metric $g_{mn}$ in 11-d SG is
related to its unimodular part $\rho_{mn}$ and  $\varphi$ as:
$g_{mn}=-{\rm e}^{-2\varphi/7}\rho_{mn}$.

By using the relation (\ref{rel}), we shall obtain the corresponding
dyonic BH solutions of type-IIA superstring and the dependence of the
corresponding ADM mass on the asymptotic values of the string coupling
$e^{\phi_{\infty}}$, the volume $e^{\bar{\sigma}_{\infty}}$ and
the corresponding unimodular parts $\bar{\rho}_{mn\,\infty}$ of the
internal metric of the 6-torus.  We shall quote the ADM mass
$M_E$ in the Einstein frame, which is related to the one in the string
frame as $M_s\equiv e^{-\phi_\infty}M_E$.

We classify the solutions according to the type of 11-d fields, {\it i.e.},
the Elfbein  $E^{(11)\,A}_M$ and the 3-form $A^{(11)}_{MNP}$, from which
the 4-d $U(1)$ gauge fields are originated.  The first set of solutions is
the one corresponding to the case of KK BH's.  They fall into the following two
sets:
\begin{itemize}
\item
{\bf Type-KNR solutions\ } Magnetic charge P associated with $\bar{B}_{\mu}$,
the KK gauge field in the RR sector, and electric charge $Q_m$ associated
with $\bar{B}^m_{\mu}$, one of six KK gauge fields in the NS-NS sector:\\
\begin{eqnarray}
e^{(\phi-\phi_{\infty})} &=& \left ({{r-{\bf P}_\infty-{\bf
Q}_{m\,\infty}} \over {r-{\bf P}_\infty}} \right )^{1\over 4}, \ \
e^{2(\bar{\sigma}-\bar{\sigma}_{\infty})} =  {{(r -{\bf P}_\infty-{\bf
Q}_{m\,\infty})^2} \over {(r -{\bf P}_\infty)^ {-1}(r-{\bf
Q}_{m\,\infty})^3}},
\nonumber \\
\bar{\rho}_{mm}/\bar{\rho}_{mm\,\infty}&=&\left ( {{r-{\bf P}_\infty}\over
{r -{\bf P}_\infty-{\bf Q}_{m\,\infty}}} \right )^{5\over 6}, \ \
\bar{\rho}_{kk}/\bar{\rho}_{kk\,\infty} = \left ({{r-{\bf P}_\infty -{\bf
Q}_{m\,\infty}} \over {r-{\bf P}_\infty}} \right )^ {1\over 6}\ \ (k \neq m),
\nonumber \\
M_E &=& |{\bf P}_\infty|+|{\bf Q}_{m\,\infty}|=
e^{\bar{\sigma}_\infty/2}|P|+e^{-\phi_\infty + \bar{\sigma}_\infty/6}
\bar{\rho}^{1\over 2}_{mm\,\infty} |Q_m|.
\label{kknr}
\end{eqnarray}
The $SO(6)/SO(5)$ rotations on this solution induces ${{6\cdot 5}\over 2} -
{{5\cdot 4}\over 2} = 5$ new magnetic charge degrees of freedom in the
gauge fields $\bar{B}_m$.
For the case electric charge $Q$ and magnetic charge $P_m$ are associated
with $B_{\mu}$ and $\bar{B}^m_{\mu}$, respectively, one can obtain
the corresponding solutions by imposing the electric-magnetic duality
transformations.
\item
{\bf Type-KNN solutions\ } Magnetic charge $P_m$ associated with
$\bar{B}^m_{\mu}$ and electric charge $Q_n$ associated with
$\bar{B}^n_{\mu}$, {\it i.e.}, both charges correspond to KK fields of
NS-NS sector:\\
\begin{eqnarray}
e^{(\phi-\phi_{\infty})} &=&
\left ( {{r-{\bf Q}_{n\,\infty}} \over  {r-{\bf P}_{m\,\infty}}}
 \right)^{1\over 4}, \ \  e^{2(\bar{\sigma}-\bar{\sigma}_{\infty})} =
{{r-{\bf P}_{m\,\infty}} \over {r-{\bf Q}_{n\,\infty}}}, \ \
\bar{\rho}_{mm}/\bar{\rho}_{m\,\infty} =  {{r-{\bf P}_{m\,\infty}-{\bf
Q}_{n\,\infty}} \over  {(r-{\bf P}_{m\,\infty})^{1\over 6}(r-{\bf
Q}_{n\,\infty})^{5\over 6}}},
\nonumber \\
\bar{\rho}_{nn}/\bar{\rho}_{nn\,\infty} &=& {{(r-{\bf P}_{m\,\infty})^
{5\over 6}(r-{\bf Q}_{n\,\infty})^{1\over 6}}\over {r-{\bf P}_{m\,\infty}
-{\bf Q}_{n\,\infty}}}, \ \
\bar{\rho}_{\ell\ell}/\bar{\rho}_{\ell\ell\,\infty} =  \left ( {{r-{\bf
Q}_{n\,\infty}} \over {r-{\bf P}_{m\,\infty}}}  \right )^{1\over 6}\ \ \
(\ell \neq m,n),
\nonumber \\
M_E &=& |{\bf P}_{m\,\infty}|+|{\bf Q}_{n\,\infty}|=
e^{-\phi_\infty + \bar{\sigma}_\infty/6} \bar{\rho}^{1\over
2}_{mm\,\infty}|P_m| +  e^{-\phi_\infty + \bar{\sigma}_\infty/6}
\bar{\rho}^{1\over 2}_{nn\,\infty}|Q_n|,
\label{kknn}
\end{eqnarray}
Upon imposing the $SO(6)/SO(4)$ rotations, one has the most general
supersymmetric 6-d KK BH's with constraint $\sum P_i Q_i = 0$.
\end{itemize}

Secondly, we have the following classes of dyonic solutions that
correspond to $U(1)$ gauge fields associated with the 11-d 3-form
field $A^{(11)}_{MNP}$:
\begin{itemize}
\item
{\bf Type-HNR solutions\ } Magnetic charge $P_m$ associated with
$\bar A_{{\mu}\,m}$, one of six 3-form fields in the NS-NS sector,
and the electric charge $Q_{mn}$ associated with $\bar{A}_{\mu\,mn}$,
one of fifteen 3-form fields in the RR sector:\\
\begin{eqnarray}
e^{(\phi-\phi_{\infty})} &=& \left ( {{r-2{\bf Q}_{mn\,\infty}} \over
{r-2{\bf P}_{m\,\infty}-2{\bf Q}_{mn\,\infty}}} \right )^{1\over 4}, \ \
e^{2(\bar{\sigma}-\bar{\sigma}_{\infty})} =  {{(r-2{\bf P}_{m\,\infty})
(r-2{\bf Q}_{mn\,\infty})} \over  {(r - 2{\bf P}_{m\,\infty}-2{\bf Q}_
{mn\,\infty})^2}},
\nonumber \\
\bar{\rho}_{mm}/\bar{\rho}_{mm\,\infty} &=& {{(r-2{\bf P}_{m\,\infty})^
{-{2\over 3}}(r-2{\bf Q}_{mn\,\infty})^{5\over 6}} \over {(r-2{\bf P}_
{m\,\infty}-2{\bf Q}_{mn\,\infty})^{1\over 6}}}, \ \
\bar{\rho}_{nn}/\bar{\rho}_{n\,\infty} = {{(r-2{\bf P}_{m\,\infty}-
2{\bf Q}_{mn\,\infty})^{5\over 6}} \over  {(r-2{\bf P}_{m\,\infty})^
{2\over 3}(r-2{\bf Q}_ {mn\,\infty})^{1\over 6}}},
\nonumber \\
\bar{\rho}_{\ell\ell}/\bar{\rho}_{\ell\ell\,\infty} &=& {{(r-2{\bf P}_
{m\,\infty})^{1\over 3}(r-2{\bf Q}_{mn\,\infty})^{-{1\over 6}}}
\over {(r-2{\bf P}_{m\,\infty} - 2{\bf Q}_{mn\,\infty})^{1\over 6}}}\ \
(\ell \neq m,n),
\nonumber \\
M_E &=& 2|{\bf P}_{m\,\infty}|+2|{\bf Q}_{mn\,\infty}|=
2e^{-\phi_\infty-\bar{\sigma}_\infty/6} \bar{\rho}^{-{1\over 2}}_
{mm\,\infty}|P_m| +  e^{\bar{\sigma}_\infty/6}\bar{\rho}^
{-{1\over 2}}_{mm\,\infty} \bar{\rho}^{-{1\over 2}}_
{nn\,\infty}|Q_{mn}|.
\label{threenr}
\end{eqnarray}
The $SO(6)/SO(4)$ rotations induce ${{6\cdot 5}\over 2} - {{4\cdot 3}
\over 2} = 9$ new charge degrees of freedom.  For the case of the
electric charge $\bar Q_m$ coming from $\bar A_{\mu\,m}$ and
magnetic charge $P_{mn}$ coming from $A_{\mu\,mn}$, the corresponding
solutions can be obtained by imposing the electric-magnetic duality
transformations.
\item
{\bf Type-HRR solutions\ }  Magnetic charge $P_{mn}$ coming from
$A_{\mu\,mn}$ and electric charge $Q_{mp}$ coming from $A_{\mu\,mp}$,
both of which are the charges of $U(1)$ fields in R-R sector:\\
\begin{eqnarray}
e^{(\phi-\phi_{\infty})}&=&1, \ \  e^{2(\bar{\sigma}-\bar{\sigma}_{\infty})}
=  {{r - 2{\bf P}_{mn\,\infty}} \over {r - 2{\bf Q}_{mp\,\infty}}},
\nonumber \\  \bar{\rho}_{mm}/\bar{\rho}_{mm\,\infty} &=&
\left ({{r-2{\bf P}_{mn\,\infty}} \over {r-2{\bf Q}_{mp\,\infty}}}\right )^
{-{2\over 3}}, \ \  \bar{\rho}_{nn}/\bar{\rho}_{nn\,\infty} =
{{(r-2{\bf P}_{mn\,\infty})^{1\over 3}(r-2{\bf Q}_{mp\,\infty})^
{2\over 3}}  \over {r-2{\bf P}_{mn\,\infty}-2{\bf Q}_{mp\,\infty}}},
\nonumber \\
\bar{\rho}_{pp}/\bar{\rho}_{p\,\infty} &=&  {{r - 2{\bf P}_{mn\,\infty}-
2{\bf Q}_{mp\,\infty}} \over  {(r-2{\bf P}_{mn\,\infty})^{2\over 3}
(r-2{\bf Q}_ {mp\,\infty})^{1\over 3}}}, \ \
\bar{\rho}_{\ell\ell} = \left ({{r-2{\bf P}_{mn\,\infty}} \over
{r-2{\bf Q}_{mp\,\infty}}} \right )^{1\over 3}\ \ \ (\ell \neq m,n,p),
\nonumber \\
M_E &=& 2|{\bf P}_{mn\,\infty}|+2|{\bf Q}_{mp\,\infty}|=
2e^{\bar{\sigma}_\infty/6}\bar{\rho}^ {-{1\over 2}}_{mm\,\infty}
\bar{\rho}^{-{1\over 2}}_{nn\,\infty} |P_{mn}| +
e^{\bar{\sigma}_\infty /6}\bar{\rho}^{-{1\over 2}}_{mm\,\infty}
\bar{\rho}^{-{1\over 2}}_{p\,\infty} |Q_{mp}|,
\label{threenn}
\end{eqnarray}
The $SO(6)/SO(2)$ transformations on this solution introduces
${{6\cdot 5}\over 2}-{{2\cdot 1}\over 2} = 14$ charge degrees of freedom
in the gauge fields $A_{\mu\,ij}$.
\item
{\bf Type-HNN solutions\ } Magnetic charge $P_m$ associated with
$\bar{A}_{\mu\,m}$ and electric charge $Q_n$ associated with
$\bar{A}_{\mu\,n}$, {\it i.e.}, both charges arise from 3-form
fields in the NS-NS sector
\footnote{Note, that for a special case with either $\bar{P}_m=0$ or
$\bar{Q}_n=0$, the result reduces to a solution first found in Ref.
\cite{BANKS}, and it is related to $H$-monopoles of heterotic string
\cite {KHURI,HARVEY}.}: \\
\begin{eqnarray} e^{(\phi-\phi_\infty)} &=&
\left ({{r-2{\bf Q}_{n\,\infty}}\over{r-2{\bf P}_{m\,\infty}}}\right )^
{1\over 4}, \ \
e^{2(\bar{\sigma}-\bar{\sigma}_\infty)} = {{r-2{\bf Q}_{n\,\infty}} \over
{r-2{\bf P}_{m\,\infty}}},
\nonumber \\
\bar{\rho}_{mm}/\bar{\rho}_{mm\,\infty} &=& {{(r-2{\bf P}_{m\,\infty})^
{1\over 6}(r-2{\bf Q}_{n\,\infty})^{5\over 6}} \over{r-2{\bf P}_{m\,\infty}
- 2{\bf Q}_{n\,\infty}}}, \ \
\bar{\rho}_{nn}/\bar{\rho}_{nn\,\infty} = {{r-2{\bf P}_{m\,\infty}
-2{\bf Q}_{n\,\infty}} \over {(r-2{\bf P}_{m\,\infty})^{5\over 6}
(r-2{\bf Q}_{n\,\infty})^{1\over 6}}},
\nonumber \\
\bar{\rho}_{\ell\ell}/\bar{\rho}_{\ell\ell\,\infty} &=&
\left ( {{r-2{\bf P}_{m\,\infty}} \over {r-2{\bf Q}_{n\,\infty}}}
\right )^{1\over 6} \ \ \ (\ell \neq m,n),
\nonumber \\
M_E &=& 2|{\bf P}_{m\,\infty}|+2|{\bf Q}_{n\,\infty}|=
e^{-\phi_\infty-\bar{\sigma}_\infty/6}
\bar{\rho}^{-{1\over 2}}_{mm\,\infty}|P_m| +
e^{-\phi_\infty-\bar{\sigma}_\infty/6} \bar{\rho}^{-{1\over 2}}_
{nn\,\infty}|Q_n|.
\label{threerr}
\end{eqnarray}
Upon imposing the $SO(6)/SO(4)$ rotations, one obtains the most general
supersymmetric 4-d 2-form BH solutions with the constraint $\sum Q_i P_i=0$.
\end{itemize}
In the expressions for the  Einstein frame ADM mass $M_E$ [string frame
ADM mass $M_s=e^{-\phi_{\infty}}M_E$] the screened charges from the RR
sector do not scale [scale as $e^{-\phi_{\infty}}$] with respect to the
asymptotic string coupling $e^{\phi_{\infty}}$, while the screened
charges from the NS-NS sector scale as $e^{-\phi_{\infty}}$ [scale as
$e^{-2\phi_{\infty}}$], in agreement with a general analysis in Ref.
\cite{WITTEN}.  Note also the following scaling dependence of screened
charges on the asymptotic volume of the 6-torus, parameterized by
$e^{\bar{\sigma}_{\infty}}$: charges, associated with the KK fields in
the RR and NS-NS sector, scale as $e^{\bar{\sigma}_{\infty}/6}$ and
$e^{\bar{\sigma}_{\infty}/2}$, respectively, while charges associated
with the 3-form fields all scale as $e^{-\bar{\sigma}_{\infty}/6}$.

We would also like to comment on the symmetry between the solutions with
gauge fields originating from the NS-NS sector, {\it i.e.}, KK gauge fields
$\bar{B}^m_{\mu}$ and the $U(1)$ gauge fields $\bar{A}_{\mu\,m}$ originated
from the 10-d 2-form $A_{\breve{\mu}\breve{\nu}}$.  That is to say, NS-NS
sector of the zero-slope limit of type-IIA string compactified on a 6-trous,
which is the same as the zero-slope limit of the corresponding heterotic
string with gauge fields in the left-moving sector turned off, has the
$SO(6,6)$ target space symmetry.  This symmetry transformations transform
the six KK gauge fields $\bar{B}^m_{\mu}$ and the six 2-form $U(1)$ gauge
fields $\bar{A}_{\mu\,m}$, along with the internal metric $\bar g_{mn}$ and
the corresponding pseudo scalar fields $\bar{A}_{mn}$, while leaving the
4-d space-time metric and the 4-d dilaton $\phi$ intact.  In fact,
a discrete subset of $SO(6)$ transformations exchanges KK gauge fields and
the 2-from $U(1)$ gauge fields in the NS-NS sector, and transforms the
internal metric $\bar{g}_{mn}$ into its inverse $(\bar{g}^{-1})^{mn}$
\cite{HET}, thus relating the Type-KNN and Type-HNN solutions.
This discrete symmetry generalizes the duality between $H$-monopole
solution and KK solution in 5-d superstring \cite{BANKS}.
Ultimately, one would like to address the full U-duality symmetry
$E_7$, which relates all the above solutions.

\section{Conclusions}

We discussed two separate classes of 4-dimensional (4-d), supersymmetric,
dyonic, spherically symmetric, black holes (BH's) in 11-d supergravity (SG)
compactified on a 7-torus.  The first class of solutions is associated
with 7 $U(1)$ gauge fields $B^i_\mu$ ($i=1,\cdots ,7$) of the Elfbein,
{\it i.e.},  Kaluza Klein (KK) BH's, and the second class of solutions
is associated with 21 $U(1)$ gauge fields $A_{\mu\,ij}$
($i=1,\cdots, 7,\ i<j$) originated from the 3-form field $A^{(11)}_{MNP}$,
{\it i.e.}, 3-form  BH's.  Within each class, we found the supersymmetric
solutions in the case where among scalar fields only those associated with
the internal metric of a 7-torus are turned on.

The general supersymmetric KK BH's are obtained by imposing global
SO(7) rotations on supersymmetric solutions with a diagonal internal
metric, and with one magnetic $P_j$ and one electic $Q_k$ charges coming
from different KK gauge fields.  The general supersymmetric 3-form BH's
are similarly obtained by imposing the SO(7) rotations on
supersymmetric solutions with a diagonal internal metric,
and with one magnetic $P_{ij}$ and one electic $Q_{ik}$ charges
associated with the corresponding two 3-form $U(1)$ gauge fields.
Both types of solutions, therefore, have constrained charge configurations.

We have also related the above solutions to the corresponding solutions
in the RR and NS-NS sectors of type-IIA superstring compactified on a
6-torus.  We related the ADM masses of these solutions to the screened
charges from the gauge fields in the RR and NS-NS sectors.  These screened
charges scale with the asymptotic string coupling in accordance with the
analysis is Refs. \cite{WITTEN,STRO2}.  In addition, the scaling of the
screened charges (of KK and 3-form $U(1)$ gauge fields in the RR and NS-NS
sectors) with the asymptotic value of the volume of the 6-torus are also
given.

The two classes of solutions, {\it i.e.}, the one arising from KK gauge
fields and the one from 3-form $U(1)$ gauge fields have the same
4-d metric.  Thus, we infer that there exists a larger T-duality
(or U-duality) transformation ($\in E_7$), which does not affect the
4-d space-time of configurations, that relates the two classes of
configurations.  In particular, we found a discrete transformation
which relates the two classes of supersymmetric solutions with a diagonal
internal metric.  Such a larger duality symmetry contains, as a subset, the
$SO(6,6)$ target space duality symmetry of the NS-NS sector of type-IIA
superstrings compactified on a 6-torus, whose discrete subset transforms
KK and 2-form $U(1)$ gauge fields in the NS-NS sector into one another.
Thus, this larger T-duality symmetry reduces to $SO(6,6)$ T-duality of
type-IIA superstring with the RR sector turned off, or equivalently, to the
$SO(6,6)$ T-duality symmetry of the heterotic superstring with gauge
fields of the left-moving sector turned off.

The ultimate goal is to find the generating solution which,
supplemented by a subset of $E_7$ transformations, would generate
{\it all} the supersymmetric BH solutions with all the scalar
fields turned on.  Subsets of $E_7$ transformations on the generating
solution with a particular choice of charge configurations would
in turn yield various classes of solutions which are discussed in this paper.
In that manner, one would arrive at a unified picture \cite{TOWN} of all
the 4-d, supersymmetric, spherically symmetric BH's in (type-IIA and
heterotic) string theory; different classes of BH's associated with
different types of $U(1)$ gauge fields are just the generating solutions
viewed in different reference frame of T-duality.

\acknowledgments
The work is supported by  U.S. DOE Grant No. DOE-EY-76-02-3071, and
the NATO collaborative research grant CGR 940870.

\end{document}